\documentclass[sigconf, authorversion]{acmart}

\AtBeginDocument{%
  }

\copyrightyear{2024} 
\acmYear{2024} 
\setcopyright{rightsretained} 
\acmConference{Preprint}
\acmBooktitle{July 2024} \acmDOI{10.1145/3652988.3673956}
\acmISBN{979-8-4007-0625-7/24/09}

\begin{document}

\title{Investigating User Perceptions of Collaborative Agenda Setting in Virtual Health Counseling Session}

\author{Mina Fallah}
\email{fallah.m@northeastern.edu}
\affiliation{%
 \institution{Northeastern University}
 \city{Boston}
 \state{Massachusetts}
 \country{USA}
}

\author{Farnaz Nouraei}
\email{nouraei.f@northeastern.edu}
\affiliation{%
 \institution{Northeastern University}
 \city{Boston}
 \state{Massachusetts}
 \country{USA}
}

\author{Hye Sun Yun}
\email{yun.hy@northeastern.edu}

\affiliation{%
 \institution{Northeastern University}
 \city{Boston}
 \state{Massachusetts}
 \country{USA}
}
\author{Timothy Bickmore}
\email{t.bickmore@northeastern.edu}

\affiliation{%
 \institution{Northeastern University}
\city{Boston}
 \state{Massachusetts}
\country{USA}
}


\renewcommand{\shortauthors}{Fallah et al.}

\begin{abstract}
Virtual health counselors offer the potential to provide users with information and counseling in complex areas such as disease management and health education. However, ensuring user engagement is challenging, particularly when the volume of information and length of counseling sessions increase. Agenda setting—--a clinical counseling technique where a patient and clinician collaboratively decide on session topics—--is an effective approach to tailoring discussions for individual patient needs and sustaining engagement. We explore the effectiveness of agenda setting in a virtual counselor system designed to counsel women for breast cancer genetic testing. In a between-subjects study, we assessed three versions of the system with varying levels of user control in the system’s agenda setting approach. We found that participants’ knowledge improved across all conditions. Although our results showed that any type of agenda setting was perceived as useful, regardless of user control, interviews revealed a preference for more collaboration and user involvement in the agenda setting process. Our study highlights the importance of using patient-centered approaches, such as tailored discussions, when using virtual counselors in healthcare.
\end{abstract}




\maketitle

\section{Introduction }
The growing prevalence of virtual agents in healthcare underscores their diverse and multifaceted role \cite{curtis2021improving} and their promise for delivering digital health interventions as counselors \cite{hudlicka2013virtual,aggarwal2023artificial}. Virtual counselors offer comprehensive information on health-related topics \cite{tudor2020conversational}, leveraging their access to vast amounts of health information. However, providing users with excessive health information in virtual counseling sessions can lead to cognitive overload, decreased attention, and ultimately, user disengagement 
\cite{arnold2023dealing}. Users have varying needs for health information, including the depth of detail they prefer, which necessitates personalized counseling approaches.

In health counseling, ``agenda setting'' is a technique in which the patient and healthcare provider work together at the start of a visit to identify and organize the main topics that need to be discussed \cite{stott1996professional}. Ideally, both parties collaboratively negotiate the topics that will be addressed, given their own priorities for the session \cite{stott1995innovation}. 
This approach establishes a framework for two-way conversation where the healthcare provider and the patient can share control of the conversation \cite{meeuwesen2007cultural}. Agenda setting also empowers patients by promoting autonomy and active involvement in their care \cite{koh2009motivational}. 


We explore the use of agenda setting in complex health counseling sessions with a Virtual Counselor (VC). As a case study, we focus on the delivery of counseling on breast cancer genetics for women.
This complex topic includes a significant amount of information for those new to these concepts (e.g., genes, interpreting test results, BRCA mutations, etc.) and represents an ideal candidate for setting an agenda in an initial educational interaction. 

To test the effects of collaborative agenda setting with users, we implement and compare three versions of the VC. In the first version, the agent collaboratively elicits the user's preferred agenda and negotiates for replacing topics with users during the conversation. The second version also elicits the user's preferred agenda but follows a fixed predetermined agenda in the conversation. In the third version, a predefined agenda is set by the agent with no elicitation of user preferences. We conduct a randomized between-subjects experiment to compare the three versions of the system. We hypothesize that higher levels of collaboration will lead to higher satisfaction with the interaction, and ultimately result in higher knowledge gain in cancer genetics.

\section{Design of a Virtual Counselor with Agenda Setting Capabilities}
\label{sec:design}

We designed a VC to represent a genetics counselor explaining breast cancer genetics to women. The agent is a 3D animated character that uses conversational nonverbal behavior---generated automatically by BEAT \cite{inproceedings}---
in synchrony with synthetic speech.
Dialogue is driven by hierarchical transition networks and template-based text generation, with user contributions made via multiple-choice menus of utterances.

Modeled on mock genetic counseling sessions videotaped at a cancer treatment institute in the United States, the agent interaction begins with a brief greeting and rapport-building social chat. Subsequently, the VC introduces the purpose of the session and presents a list of breast cancer topics to be discussed. The user is informed that only some topics can be covered in the current session. The VC then initiates the agenda setting protocol. After agenda setting, the VC introduces each topic in turn. 

The agenda setting between the VC and the user includes three key components: 1) Brief descriptions of each topic, 2) Elicitation of topics prioritized by users, and 3) Topic negotiation, where the agent proposes to swap one of the user's prioritized topics with an alternative one. If the user accepts this change, the VC adjusts the agenda accordingly. 

\begin{figure}
    \centering
    \includegraphics[width=0.45\textwidth]{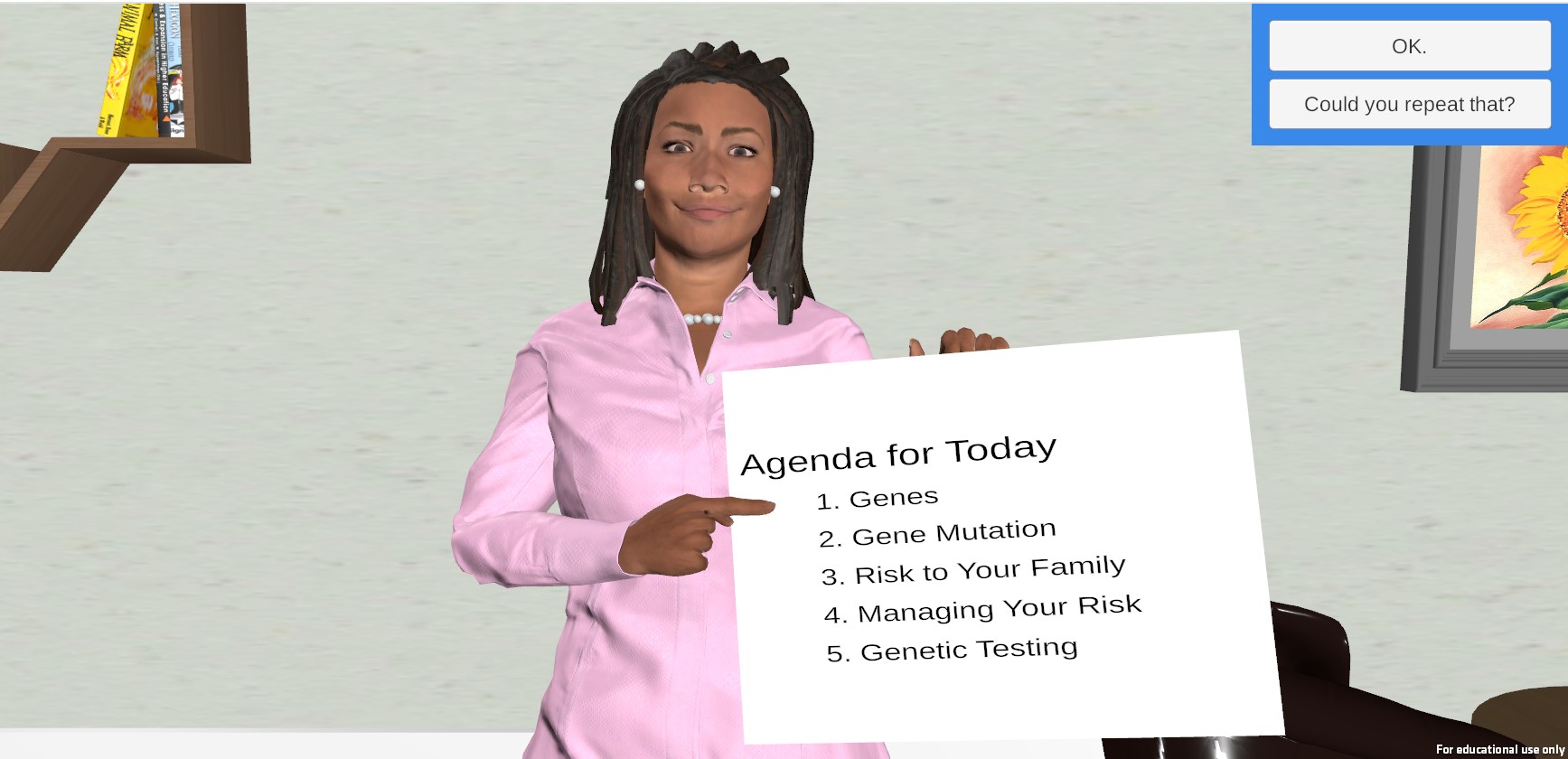}
    \caption{Virtual Counselor navigates a 20-minute session according to an agenda.}
    \label{fig:enter-label}
\end{figure}

The 20-minute counseling session starts with the agent presenting the agenda (Figure \ref{fig:enter-label}) and pointing at the first topic before delving into it.
As the session progresses, the agent re-displays the agenda at the start of each topic before proceeding to the next topic.
As the session concludes, the agent shows the agenda again and 
reviews the topics covered during the interaction.

\section{Comparative Acceptance Study}

To assess user perceptions of different levels of user involvement in the agenda setting process and its impact on satisfaction, engagement, and knowledge gain, we conducted a three-arm randomized between-subjects experiment in which the degree of agenda setting was manipulated. The three conditions were:\\
    \textbf{Collaborative Agenda Setting (COLLABORATIVE):} This condition includes the full agenda setting protocol described in \autoref{sec:design}.\\
    \textbf{Perceived Collaboration (PERCEIVED):}  In this condition, users are asked to negotiate an agenda but the agent dose not follow the negotiated agenda, instead presenting a predefined, fixed list \textit{}of topics as the final agenda. 
    The VC tells the user that it has created an agenda based on their interests.\\
    \textbf{No collaboration (CONTROL):} In the third condition, the agent presents a predefined, fixed agenda without eliciting user preferences.


\subsection{Study Details}

We recruited healthy English-speaking females aged 18-45, using online fliers and social media platforms for recruiting. This study was approved by our institution’s IRB, and participants were compensated for their time.

The study evaluated satisfaction with the VC interaction through a single self-report scale item (from 1=``Not at all satisfied'' to 7=``Very satisfied''). Perception of agenda setting was assessed using a 19-item scale (from 1=``Somewhat'' to 7=``Extremely'') based on \cite{phdthesis}. The participants' knowledge of breast cancer genetics was assessed immediately before and after interacting with the virtual counselor using an 11-item true-false quiz \cite{lerman1996brca1,lerman1997controlled}.  

Participants completed the study remotely on Zoom. They were randomly assigned to one of the three study conditions and interacted with the virtual counselor for approximately 20 minutes. They also completed pre and post-counseling questionnaires. Finally, a semi-structured interview was conducted, focusing on system and agent impressions, sense of control, tailoring, collaboration, and barriers to engagement.

\section{Results}
Thirty-one healthy English-speaking females ages 18-45 years old, 54.8\% Asian, 32.2\% Caucasian, 9.6\% Middle Eastern, and 3.2\% Hispanic participated in the study. All participants had completed high school or equivalent. Participants were randomly allocated to one of three conditions: COLLABORATIVE (N=11), PERCEIVED (N=10), and CONTROL (N=10).

\subsection{Quantitative Results}

Overall, participants in all conditions reported high levels of satisfaction with the experience, scoring an average of 5.1 (SD=1.6), significantly higher than a neutral score of 4.0 (Wilcoxon signed rank Z = -3.1, p<.01). That said, there were no significant differences between the three study conditions. Perceptions of the agent's agenda setting were positive across all conditions, averaging 5.59 (SD=0.76), and were significantly above the neutral score of 4.0 on these scales (single sample t(30)=33.7, p<.001). However, there were no significant differences between study conditions, F(2,28)=1.07, n.s. 
Participants in all conditions demonstrated significant increases in breast cancer genetics knowledge during their interaction with the VC, from 7.35 (SD=1.82) before the interaction to 8.48 (1.15) after (paired t(30)=4.32, p<.001). However, there were no significant differences between study conditions, F(2,28)=.52, n.s.

\subsection{Qualitative Results}
A total of 310 minutes of interviews were conducted, and transcripts were analyzed using in-vivo coding for thematic analysis.  
\subsubsection{General impressions of the system} The agent system was perceived as interactive, engaging, and easy to navigate. Most participants liked the agent and mentioned wanting to talk with it again, but some reported feeling awkward at the beginning of the interaction. The topics covered by all versions of the system were deemed important and relevant, and participants appreciated the information as well as the structured approach of the agent in teaching the material. 
\subsubsection{The level of collaboration was noticed} All participants in the CONTROL condition felt that they shared a small level of involvement in the conversation, and the topics felt \textbf{pre-determined}: ``\textit{I didn't really have control. She was just speaking about whatever topic she wanted to}'' [P9]. Another participant highlighted that the given information felt completely \textbf{programmed} and not tailored: ``\textit{I'm not sure if I'm going to get any more information than what has been programmed into her already}'' [P29]. This lack of involvement led to a belief that the agent was not collaborative enough, leaving most participants with an unmet desire to choose their own topics.

On the other hand, those in the COLLABORATIVE condition felt a high level of collaboration in the discussion, and deemed the agent as \textbf{flexible}. Specifically, participants generally enjoyed picking their own topics out of the provided list, and negotiating with the agent over replacements: ``\textit{she offered something different from what I said, I think she was very flexible...I said, I'd like to stick my agenda, and she went through with that}'' [P10]. 
Respecting participant preferences contributed to how collaborative the conversation was perceived to be.

In the PERCEIVED condition, some participants did not notice that the agenda did not incorporate their preferences. 
However, most participants were able to notice this:
"\textit{There was one point where I think...she discussed, like, what is a gene, even though I didn't select that one}", and were mostly dissatisfied with it: "\textit{One negative point I noticed...was that the agenda that I chose was different from the agenda that she was explaining to me}" [P26]. That said, these participants appreciated 
being offered topic \textbf{options}  to choose from, and felt it "\textit{really helped}" [P18] even when the options they chose were not covered in the session.

\subsubsection{Having more control over agenda was preferred} For most participants, choosing their own topics was important, as it made them feel more involved in the discussion and more \textbf{engaged} when listening to the materials. This caused participants in the COLLABORATIVE and PERCEIVED groups to develop a sense of engagement and deem the system as \textbf{interactive}: "\textit{it gave me a sense of having some control over what was shared that had me more engaged, feeling like it was my area of interest}" [P8], while these results were not found in the CONTROL group. Furthermore, participants emphasized the importance of tailoring the discussion according to their \textbf{prior knowledge}: "\textit{I know a little bit about genes and mutations, so I would rather skip that part}" [P16]. Prior knowledge on the topics  may thus be an important matter to consider for more efficient and engaging conversations with VCs. Collaborating with the agent and participating actively in the conversation was also brought up as a matter of developing a working alliance and \textbf{effective communication}, such that participants in the CONTROL group indicated the need for more of a back and forth in conversation: "\textit{If, I [could] tell her my topics and [it is] two way communication, it would be great}" [P9].

\subsubsection{The agent was trusted with topic choices, but clear explanations were preferred}
7/10 participants in the COLLABORATIVE condition accepted the topic changes negotiated by the agent, with some highlighting the \textbf{expert} role of the agent: "\textit{I picked some things that she thought were a little less important. So then she asked me if it was okay, if we talked about these things instead? And I said, Yes, because they're important. And I'd like to learn more}" [P5]. However, a couple of participants reported tensions during or after negotiations with the agent: "\textit{I felt pressured to agree with [her recommendation]}" [P13]. For one participant, a \textbf{lack of clarity} in the reason for changing their agenda led to dissatisfaction: "\textit{the one thing that struck me was that it removed one of the things that I wanted to talk about...I just picked something...[she] said I couldn't go into that. So like, what's up with my choice?}" [P22]. 

\subsubsection{Any type of agenda (structure) was deemed helpful}
In all conditions, participants appreciated the \textbf{structured} approach of the agent in delivering the educational material. This was the case across conditions, as they all followed either a predetermined or collaboratively set agenda: "\textit{it's very clear as she concluded our topic that's the end of this conversation, as she talked about every topic and concluded, and then we moved on to the next topic, which is great}" [P23]. Many participants also appreciated the initial, brief explanations of topic titles that the agent would optionally talk about, which helped them feel more prepared for listening to the topics (in all conditions) and better able to select the topics they were interested in (in COLLABORATIVE and PERCEIVED). 

\section{Discussion}

Our study examined perceptions of agenda setting with a virtual counselor in the context of a breast cancer genetics counseling session. Our quantitative findings indicated that all participants were satisfied with the experience, felt that the use of an agenda (whether collaborative or non-collaborative) was satisfactory, and significantly increased their knowledge of breast cancer genetics. However, we did not find statistically significant differences between the three versions, thus our hypothesis was not supported quantitatively. 

That said, our qualitative findings provide insights into user attitudes towards actively sharing control of the conversation with the agent. 
Whether or not participants were involved in setting the agenda, the mere existence of an explicit agenda provided a structure to the conversation, in line with prior work \cite{stott1995innovation}. Previous work in virtual agents have also shown that structured discussions increase the quality of educational interactions with these agents \cite{murali2022training}. We also found that participants who were engaged in collaborative agenda setting perceived the virtual counseling sessions as more engaging and interactive, resulting in an experience tailored to their interests. This aligns with prior research indicating that agenda-setting fosters active engagement and partnership by prioritizing topics based on patients' needs and preferences and enhancing patient autonomy consistently throughout the interaction \cite{koh2009motivational}.
  
Our findings also revealed that participants usually noticed deviations from their chosen agenda, especially when the VC presented an agenda significantly different from their stated preferences. In the PERCEIVED condition, participants reported a lack of control over the conversation, leading to dissatisfaction with the interaction. That said, the mere elicitation of preferred topics may have positively impacted engagement \cite{ilin2022role}.

Our study has several limitations, beyond the small convenience sample. A single 20-minute counseling session may not sufficiently capture the dynamic interplay and evolving priorities characteristic of agenda setting, which often unfolds across multiple interactions. Future work should test the effects of agenda setting for multiple counseling sessions, and whether tailoring educational interactions to users' prior domain knowledge can help increase satisfaction and knowledge gain.

Our present work establishes the feasibility and importance of collaboratively establishing an agenda with users in virtual counseling settings. 


\bibliographystyle{ACM-Reference-Format}
\bibliography{sample-base}




 




\end{document}